# Pulse-shaping in the interaction of an elliptically-polarized laser and magnetized-plasma

A. A. Molavi Choobini[1*], F. M. Aghamir[1], S. S. Ghaffari-Oskooei[2].

[1]Dept. of Physics, University of Tehran, Tehran 14399-55961, Iran,

[2]Department of Atomic and Molecular Physics, Faculty of Physics, Alzahra University, Tehran, Iran.

**Abstract:**

Pulse shaping provides a significant level of control and precision when optimizing laser-plasma interactions. Pulse shaping enables precise control and manipulation, resulting in enhanced energy deposition, optimized particle acceleration, controlled polarization, and exploitation of resonant effects. The present study investigates the interaction of structured light with magnetized plasma, considering various spatial profiles and polarization states. This phenomenon involves modification of the temporal and spatial characteristics of the laser pulse due to the presence of the magnetized plasma. The discussion reveals how the electric field and electron velocity evolve within the plasma both spatially and temporally. Factors such as absorption, dispersion, collisions, and scattering are taken into account to understand how they influence the evolution of the pulse. The effects of electron density, external magnetic fields, relativistic velocities, and polarization states on pulse compression are examined. The spatial laser profile impact on pulse-shaping and plasma channel formation is also discussed. This exploration sheds light on the intricate interplays and potential pulse-shaping applications in laser-plasma interactions.

## I. Introduction

The exploration of generating short laser pulses from optical counterparts stands as a challenging frontier within the realm of nonlinear optics [1, 2]. The non-linear progression of highly intense laser beams through a transparent medium and the interaction between intense pulses and matter produces numerous nonlinear physical phenomena [3]. Short pulses find utility across a wide spectrum encompassing both scientific and applied research, with a specific emphasis on exploring electron dynamics using precise time-domain spectroscopy techniques. These concise pulses offer a versatile toolkit for exploring rapid electronic processes and dissecting the intricacies of laser schemes driving particle acceleration, wakefield accelerators, and the fascinating phenomenon of high harmonic generation [4, 5].

In many of these applications, shaping or compression of laser pulses is of paramount importance, resulting spatial and temporal compression of light while propagating at heightened intensity. Over the past decade, there has been a noteworthy surge in research focused on compressing and shaping laser pulses to achieve exceptionally high peak powers [6-8]. As a laser pulse traverses a medium, it undergoes a phenomenon known as self-phase modulation (SPM), shaping and compression, leading to the expansion of its spectral profile. In response to these interactions, the spectral bandwidth of laser pulses can be effectively compressed through the strategic application of prism or grating pairs. Furthermore, the interaction between short laser pulses and plasma introduces the intriguing concept of relativistic nonlinearities, enhancing the prospects of laser pulse compression [9, 10]. Currently, multiple approaches are being developed to harness nonlinear processes for laser pulse compression, encompassing both plasma and nonlinear crystals. For this context, nonlinear characteristics of plasma offer an added advantage compared to waveguides, gratings, and nonlinear crystals. Plasma can endure significantly higher peak-intensity laser pulse exposure than diffraction gratings, optical crystals, and nonlinear optical materials. Furthermore, the interaction of a short pulse with plasma induces relativistic nonlinear effects, making it suitable for laser pulse shaping and compression. This phenomenon holds promise for pushing the boundaries of laser technology, thereby expanding our understanding of light-matter interactions and driving innovation within the realm of nonlinear optics, and investigations are ongoing on developing innovative ways for this field [11, 12].

S. Kumar and colleagues investigated self-compression improvement via co-propagating laser pulses in plasma [13]. They examined modified nonlinear Schrödinger equations for low and high-intensity laser pulses and found that the compression of high-intensity pulses significantly depends on the combined intensity of lasers. The research by T. C. Wilson et al. delved into the theoretical and numerical examination of three-dimensional compression dynamics of highly intense ultrashort laser pulses in plasma [14]. Their simulations showed the possibility of achieving spherical compression of the laser pulse to a size approximately equal to the laser wavelength, referred to as the lambda-cubic regime and the collapse can occur multiple times during the laser pulse propagation. P. Panagiotopoulos and his team investigated a direct approach for generating high-power, ultrashort laser pulses at 10 μm by self-compression for the next-generation of relativistic phenomena [15]. Through the interplay of self-phase modulation and anomalous dispersion, they achieved a significant compression factor of approximately 3.5 times. This compression was followed by the emergence of filamentation near the cell. X. Gao and B. Shim considered the self-focusing and self-compression of intense pulses via ionization-induced spatiotemporal reshaping [16]. They found that the pulse undergoes self-focusing and self-compression through ionization-induced reshaping, resulting in a manyfold increase in laser intensity. M. R. Edwards and Pierre Michel described the design of a compact high-power laser system that uses plasma transmission grating, with currently achievable parameter, for chirped pulse amplification [17]. Their simulations revealed that the meter-scale final grating for a 10-PW laser could be replaced with a 1.5-mm-diameter plasma grating, allowing compression and providing a path toward compact multi-petawatt laser system. In an experimental study, M. Jaberi and his colleagues investigated how interaction geometry parameters impact the stimulated Brillouin scattering process within a generator cell, aiming to optimize compression [18]. Their findings showed that manipulating the lens distance and input energy to the cell allows for precise control over pulse compression ratio, enabling the attainment of maximum pulse compression. N. Gupta and his team conducted a theoretical exploration into the self-compression of a laser pulse possessing a q-Gaussian spatial irradiance profile [19]. Employing a variational approach rooted in nonlinear Lagrangian formulation, they analyzed how laser-plasma parameters influence the laser pulse propagation dynamics.

In the present study, a pioneering mechanism is presented for the investigation of pulse-shaping in the context of elliptically polarized laser and magnetized plasma interaction with consideration of various effects. One of the pivotal aspects explored in this study is the effect of a magnetic field within the plasma environment. The presence of a magnetic field in the plasma environment significantly alters the dynamics of self-focusing phenomena, by intricately modulating dispersion properties and affecting plasma responses, the magnetic field engenders pulse-shaping behaviors that markedly differ from those observed in non-magnetized plasma scenarios. Elliptical polarization, another remarkable facet of this study, introduces a distinctive and intriguing dimension to pulse compression in laser-plasma interactions. The effect of elliptical polarization on pulse-shaping can lead to modified compression dynamics and enhanced control to shape the compressed pulse temporal and spectral properties. Furthermore, the analysis evaluates the effect of the laser pulse profile, as well as the electron density on the dynamics of the pulse-shaping process, recognizing their pivotal role in shaping the overall pulse compression mechanism. A well-defined pulse profile can lead to a tightly focused and controlled output. Dense regions of plasma slow down the pulse, causing localized compression, while lower-density regions allow the pulse to propagate faster, influencing the overall pulse shape. The paper is arranged as follows: in section II, the theoretical model for numerical simulations is considered. The next section presents our results and the discussions therein. Conclusions are drawn in section IV.

## II. Analytical Model

An elliptically-polarized laser, propagating in a magnetized plasma, is considered. It is assumed that its electrical field is given by the following expression:

$$\vec{E}_L = E_x(z,t)\hat{e}_x + E_z(z,t)\hat{e}_z = \frac{1}{2}[E_{0x}\hat{e}_x + iE_{0z}\hat{e}_z]\mathrm{P}(Z,T) + c.c. \quad (1)$$

here $E_{0x}$ and $E_{0z}$ are the amplitudes of the electrical fields of laser beam in the $x$ and $z$ directions. The function of $\mathrm{P}(Z,T)$ is defined as the profile of laser pulse, where $Z(z)$ and $T(t)$ are spatial and temporal parts of its profile, respectively. A spatial laser profile function $Z(z) = \mathrm{Cosh}\left(\frac{zb}{z_0}\right)\exp\left[-\left(\frac{z}{z_0}\right)^q\right]e^{ikz}$, is considered to describe how different laser pulse profiles interact with plasma electrons, where $k$ represents the laser field wave number, $q$ (the index number) and $b$ (the skew parameter) which stipulate various laser profiles, and $z_0$ is the width of the laser field. By adjusting these parameters, various laser profiles of all kinds can be realized. The temporal part of the laser profile is assumed to be a Gaussian and varying as $e^{-i\omega t}$, where $\omega$ represents the laser frequency. In an un-magnetized plasma, the propagation of waves is influenced by their polarization, meaning that the wave vector remains constant for a given laser frequency. However, the introduction of a magnetic field plasma environment alters the dynamics significantly, the plasma electrons begin to exhibit Larmor motion, circling around the magnetic field lines. Specifically, when a static magnetic field is applied along the axis of laser propagation, it creates an anisotropic environment within the plasma. Thus, plasma is considered to be in a DC magnetic field $\vec{B}_0 = B_0 \hat{e}_y$.

Numerical solution of the momentum of the plasma electrons and electromagnetic wave equation is the basis of the present study. The insertion of the electric field of the laser pulse in wave equation yields:

$$\vec{\nabla}(\vec{\nabla}.\vec{E}) - \nabla^2 \vec{E} + \frac{1}{c^2}\frac{\partial^2 \vec{E}}{\partial t^2} = -\frac{4\pi}{c^2}\frac{\partial \vec{J}}{\partial t} \qquad (2)$$

where $\vec{J} = -ne\vec{v}$ represents the current density, $n$ is the density of electrons and $\vec{v}$ is electron plasma velocity. The consideration of $E_x(z,t)$, $E_z(z,t)$ and application of partial derivatives lead to the following equations:

$$-A(z)\frac{\partial^2 E_z(z,t)}{\partial z^2} + (B(z) - 2ik)\frac{\partial E_z(z,t)}{\partial z} - \frac{2ik}{v_g}\frac{\partial E_z(z,t)}{\partial t} - \frac{\omega^2}{c^2}E_z(z,t) = \frac{-4\pi i \omega n}{c^2}v_z \qquad (3)$$

and

$$-A(z)\frac{\partial^2 E_x(z,t)}{\partial z^2} + (B(z) - 2ik)\frac{\partial E_x(z,t)}{\partial z} - \frac{2ik}{v_g}\frac{\partial E_x(z,t)}{\partial t} + \left(k^2 - \frac{\omega^2}{c^2}\right)E_x(z,t) = \frac{-4\pi i \omega n}{c^2}v_x \qquad (4)$$

here $A(z) = \partial^2 Z(z)/\partial z^2$, $B(z) = \partial Z(z)/\partial z$, $v_x$ and $v_z$ are electrons velocities in the $x$ and $z$ directions respectively, and $v_g = c^2 k/\omega$ is the group velocity of laser pulse in the presence of the magnetic field. For the assessment of electron velocity, the relativistic momentum equation for electrons can be used to evaluate pulse-shaping dynamics:

$$\frac{d\vec{p}}{dt} = -e\left[\vec{E} + \frac{\vec{v}}{c} \times \vec{B}\right] \qquad (5)$$

here $\vec{p} = \gamma m_e \vec{v}$, $m_e$ is the electron mass and $\gamma = \sqrt{1 + |\vec{p}|^2/m_e^2 c^2}$ is Lorentz factor. The magnetic field $\vec{B} = \vec{B}_L + \vec{B}_0$ includes two portions; a part due to laser pulses and an external homogeneous field. By substituting the electric and magnetic fields in Eq. (5), two coupled equations for components of electron velocity are derived:

$$\frac{d(\gamma m v_x)}{dt} = -eE_x + \frac{e}{c}v_z B_0 \qquad (6)$$

$$\frac{d(\gamma m v_z)}{dt} = -eE_z - \frac{e}{c}v_x B_0 \qquad (7)$$

By linearizing the above equations and performing the required algebraic operations, the relationship between the components of the electric field in the X-mode will be as follows:

$$E_z(z,t) = -\frac{i\gamma^2 \omega}{\omega_p^2 \omega_c}[(\omega^2 - c^2 k^2)\left(1 - \frac{\omega_c^2}{\gamma^2 \omega^2}\right) - \frac{\omega_p^2}{\gamma}]E_x(z,t) \qquad (8)$$

where $\omega_c$ and $\omega_p$ are cyclotron and plasma frequency, respectively. The numerical investigation of the properties of the laser pulse-shaping including variations of the electric fields and electron velocity are evaluated through solving Eqs. (3, 4, 6 and 7) simultaneously.

## III. Results and Discussion

Pulse shaping as an indispensable method provides meticulous control over laser pulses' temporal and spectral features. By matching the pulse shape to resonate with specific plasma oscillations or modes, one can enhance the efficiency of energy transfer and particle acceleration. Tailoring shaping the pulse, one can optimize the parameters of the laser, ensuring it interacts with the plasma in a manner that produces desired effects, such as controlled energy deposition and particle acceleration. The focused energy delivery minimizes wasteful energy loss and ensures that the available energy is utilized efficiently in the interaction process. A properly shaped pulse can establish a stable and controlled acceleration gradient within the plasma. This stable gradient leads to particles accelerated uniformly and efficiently, resulting in the production of higher-energy particles with greater precision. The present analysis presents the pulse-shaping techniques in the interaction of laser pulses with magnetized plasma having different profiles and polarization states under varying conditions. Figure 1 illustrates the laser pulse's interaction with magnetized plasma. In panel (a), a schematic showcases the process where the input laser pulse undergoes stretching and compression during the interaction. Panel (b) displays the amplitude variations of the compressed electric field generated in uniform plasma concerning the spatial coordinate. These insights, derived from computational models, are then rigorously compared with experimental outcomes, forming a robust feedback loop that refines our understanding and applications of pulse-shaping techniques. The following experimental parameters listed in references [20, 21] were considered for simulation runs; laser wavelength $\lambda = 800$ nm, laser pulse duration $\tau_L = 100$fs at full width at half maximum, and peak incipient laser pulse intensity at 42 mJ is taken as $I_0 = 5 \times 10^{16}$ W/cm$^2$.

Figure 2 shows a 3D plot depicting a comprehensive visual representation of the dynamic behavior of the normalized electric field as a function of the normalized z coordinate over plasma length and as a function of normalized time over pulse duration for a Gaussian profile. Upon close examination of the plot, it becomes evident how the electric field evolves both spatially and temporally within the plasma. From the plot, it can be observed that the electric field initially starts with a high intensity at the beginning of the pulse and gradually compresses while the shape of the pulse changes as it propagates through the plasma. Several key factors contribute to this observed pulse shaping over the plasma length. The absorption mechanisms within the plasma lead to alterations in the electric field intensity, the dispersion phenomena, where the laser pulse components travel at different velocities. The scattering and collisions between particles within the plasma medium lead to modifications in the electric field's behavior. In validation of these results, Min Sup Hur et. al verified experimentally the compression of electric field by plasma density gradient [22]. The 3D plot illustrating the variation of normalized electron velocity as a function of normalized z coordinate over plasma length and normalized time over pulse duration, divided into seven-time intervals, for a Gaussian profile is drawn in Figure 3. The plot shows how the electron velocity changes both spatially and temporally within the plasma. At the beginning of the interaction, the plasma electrons receive momentum and energy from the laser pulses, and the size, as well as number of oscillations gradually increase. Over the time, due to collisions, electrons energy decreases, leading to a reduction in the number of oscillations and their size. In addition, Figure 3 reveals that the electron velocity experiences oscillations near certain time intervals, indicating the influence of cyclotron resonance. The resonance occurs when the frequency of the laser pulse matches the cyclotron frequency of the electrons. The enhanced relativistic nonlinearity near the gyro-resonance can lead to significant changes in the electron velocity.

Plasma-based compression techniques utilizing plasma gratings or plasma channels offer a means to induce precise phase modulation and achieve compression effects. These techniques leverage the unique properties of plasmas to control the phase of laser pulses, ultimately leading to pulse compression. In the case of plasma gratings, plasma structures are ingeniously engineered to serve as "gratings" for modulating the phase of the incident laser pulse. This modulation relies on the careful tailoring of electron density and plasma frequency within the plasma structure to achieve the desired compression effects. Figure 4 depicts the variations of the magnitude of normalized electric field as a function of normalized z coordinate over plasma length for different electron densities and Gaussian profile. As the figure indicates, electron density variations in the plasma create regions of different refractive indices. The plasma frequency determines the spacing of these regions and influences the refractive index of the plasma, which in turn affects the phase velocity of the laser pulse and its modulation. The self-phase modulation (SPM) occurs when the intensity-dependent refractive index induces a frequency shift in different parts of the pulse. By exercising precise control over the electron density within the plasma structure, one can govern the degree of phase modulation, thereby enabling pulse-shaping and compression through constructive interference. In other words, the plasma frequency introduces dispersion which refers to the dependence of a material refractive index on the frequency of the laser pulse. Different frequency components of pulse travel at different speeds due to the dispersion relation and the resulting spatial delays or spatial phase difference between components leading to pulse compression, affecting pulse-shaping.

The effects of an external magnetic field on pulse compression in laser-plasma interactions are the result of intricate interplays between electromagnetic forces, plasma dynamics, and the characteristics of the laser pulse. In a high-intensity laser pulse interacting with a plasma, the laser can undergo pulse-shaping due to the ponderomotive force. This force can cause electrons to oscillate, forming density variations in the plasma and causing the density profiles and refractive properties of plasma altered, leading to changes in the laser spatial and temporal characteristics. In addition, the Lorentz force acting on charged particles due to the external magnetic field can compress the plasma. Charged particles experience a magnetic force perpendicular to their velocity and the magnetic field direction. This could affect the spectral components of the laser pulse, potentially leading to different compression dynamics. Therefore, an external magnetic field can alter the trajectories of plasma electrons, affecting their response to the ponderomotive force and consequently influencing pulse-shaping. Effect of various external magnetic field on the magnitude of the normalized electric field as a function of normalized z coordinate over plasma length for Gaussian profile is depicted in Fig 5. According to the figure, the presence of a static magnetic field in the plasma affects the speed of electrons in the plasma through cyclotron resonance, especially in the elliptically polarized state, and changes the direction of oscillation. This leads to enhanced relativistic nonlinearity near the gyro-resonance, significantly affecting the effective dielectric constant and other propagation parameters of the X-mode, In turn, it has a direct impact on pulse-shaping and compression. Furthermore, the X-mode has the advantage of being able to propagate even in dense plasmas, as long as the laser frequency is lower than the cyclotron frequency.

Figure 6 shows the plot of the normalized electric field as a function of normalized z coordinate over plasma length for different $\beta = v/c$ parameter and Gaussian profile. From the Figure, it is evident that the effects of relativistic velocities in the interaction between a high-intensity laser and plasma can have significant implications for pulse compression and the overall interaction dynamics. As particles approach relativistic velocities, their kinetic energy increases drastically, leading to a higher effective mass due to the Lorentz factor ($\gamma$). With the increased effective mass, the ponderomotive force becomes stronger, influencing plasma dynamics and wave generation more significantly. At relativistic velocities, electrons experience significant Lorentz contraction and mass increase, as a result, the stronger deflections in the presence of the laser electric field occur. The enhanced electron motion leads to more pronounced nonlinear behavior, including self-focusing and plasma wave generation. The relativistic electron dynamics can contribute to enhanced compression or alter the conditions under which certain compression mechanisms operate. In addition, relativistic velocities introduce time dilation, where observers moving at different velocities perceive different times. This time dilation leads to frequency shifts in the doppler effect. As plasma particles move at relativistic velocities, their emissions and scattering of light experience these frequency shifts, which can alter the observed spectral characteristics of the pulse.

The polarization direction of the electric field determines how the electromagnetic wave oscillates in space. Different polarization states lead to different behaviors in the plasma, affecting compression dynamics. Figure 7 displays the variations of the magnitude of the normalized electric field as a function of normalized z coordinate over plasma length for different polarization states of laser pulse. The figure shows that for all polarization states, nonlinear plasma responses are affected by the polarization, altering the phase-matching conditions. This can lead to different patterns of electron motion which influence pulse compression. Circular polarization introduces a rotational component to the electron trajectories due to the rotating electric field. This rotation leads to a phase modulation pattern across the pulse, affecting how different pulse components interfere. This outcome for the normalized electric field of pulse-shaping in plasma is confirmed by the experimental findings of Jihoon Kim et al. [21]. In elliptical polarization ($E_{0x} \neq E_{0z}$) leads to stronger ponderomotive forces acting on electrons due to the enhancement of the one of components of the electric field compared to both right-handed ($E_{0x} = E_{0z}$) and left-handed ($E_{0x} = -E_{0z}$) circularly polarized electric fields. The motion of electrons in the plasma, induced by these forces, contributes to pulse-shaping, self-focusing, and phase modulation, causing more wave compression. Therefore, the effects of electric field polarization on pulse compression arise from the intricate interplay between the polarization direction, the response of charged particles in plasma, and interactions driven by the laser electromagnetic field. By manipulating the polarization state, one can control these interactions and tailor the compression process for specific applications, taking advantage of the polarization-dependent behaviors of plasma and laser pulse-shaping.

Figure 8 depicts the effect of various spatial laser profiles on the magnitude of the normalized electric field as a function of normalized z coordinate over plasma length. The spatial profile of a laser pulse, often referred to as its beam profile, plays a significant role in the interaction between the laser and plasma, particularly in the context of pulse compression. Nonlinear plasma responses and absorption of it depend on the local intensity of the laser pulse and the spatial profile determines how the intensity is distributed across the beam. A nonuniform profile can lead to varying levels of nonlinear interaction and absorption within the plasma, influencing the pulse compression process. The figure indicates that the ring-shaped profile ($q = 0, b = 1$) has a significantly higher amplitude than the supper-Gaussian profile ($q = 2, b = 0$). The cosh-Gaussian profile ($q = 0, b = 2$) has also a stronger electric field than the others due to better phase matching between the electron velocity and the phase velocity of the laser pulse. When these velocities are well-matched, the laser pulse can efficiently transfer energy to the plasma electrons, which can result in a stronger induced electron density and, in some cases, stronger electric fields. In plasma, the intensity of the spatial laser pulse profile affects the plasma electron density due to the ponderomotive force. Figure 9 shows the variations in the magnitude of the normalized intensity as a function of normalized retarded time throughout the duration of the laser pulse for different laser profiles. This indicates the efficiency of pulse shaping, allowing for a direct comparison between different laser profiles and their effects on the pulse-shaping process. With an increase in the skew parameter of the laser pulse and the use of the cosh-Gaussian profile compared to super-Gaussian, the ponderomotive force of laser pulses is enhanced, leading to an augmentation in the compression of pulse and the efficiency of the pulse. Increasing the force leads to more pronounced self-focusing as the peak intensity can create a higher electron density region, enhancing the focusing effect. Various profiles induce more controlled plasma channels, affecting the propagation and compression of the pulse due to variable index number and the skew parameter. Therefore, different spatial profiles can introduce phase variations across the pulse due to differences in the propagation of the various beam components. These phase variations can interact with plasma and lead to spatial phase modulation. This modulation can affect the compression process, as the phase of the pulse components influences their interference patterns and temporal characteristics. Furthermore, controlled plasma channels can be used to guide and compress the laser pulse. A specific profile may match better with the geometry of the channel, optimizing the pulse-shaping process. The intensity distribution can also influence the power deposition along the channel, affecting compression efficiency.

## IV. Conclusions

The understanding of pulse shaping in laser-plasma interactions is crucial for optimizing compression techniques and improving the efficiency of these interactions. By analyzing the behavior of electric fields and electron velocities within a plasma, valuable insights can be gained. This study delves into the mechanism of pulse-shaping in the interaction of an elliptically-polarized laser and magnetized plasma. Pulse-shaping within the plasma occurs due to factors like absorption, dispersion, collisions, and scattering, affecting the evolution of the electric field and electron velocity over time. The presented scheme provides adjustable factors to change the pulse-shaping characteristics. Electron density modulation in the plasma, the presence of an external magnetic field, and relativistic velocities of plasma electrons each significantly impact compression dynamics. The variations of the normalized electric fields for various polarization states were examined, and it was found that the polarization state of the laser has a profound effect on pulse-shaping, leading to varied electron motion patterns and phase modulation effects. The spatial profile of the laser pulse influences compression by determining the local intensity distribution, inducing phase variations, and affecting the evolution of plasma channels. Additionally, the variations in electron velocity can be investigated to identify time intervals where cyclotron resonance occurs. This phenomenon can be exploited to enhance particle acceleration or manipulate plasma waves, leading to more efficient particle acceleration or generation of specific plasma wave structures for various applications. These findings collectively demonstrate the intricate interplay between pulse shaping, plasma properties, and external factors in laser-plasma interactions.

## Acknowledgment

This research did not receive any specific grant from funding agencies in the public, commercial, or not-for-profit sectors.

## List of Figures & Captions

**Fig. 1.** (a) Schematic of the interaction of laser pulse with magnetized plasma, an input laser pulse is stretched and compressed after the interaction, (b) Variations of the amplitude of the compressed electric field generated in the uniform plasma as a function of the spatial coordinate.

**Fig. 2.** The 3D plot of variation of normalized electric fields as a function of normalized z coordinate over plasma length and as a function of normalized time over pulse duration for Gaussian profile, $n = 3.5 \times 10^{16} \, W/c^2$, $\beta = 0.01$, $B_0 = 100T$, and elliptical polarization state.

**Fig. 3.** The 3D plot of variation of normalized electron velocity as a function of normalized z coordinate over plasma length and as a function of normalized time over pulse duration in 8 divided time intervals and for Gaussian profile, $n = 3.5 \times 10^{16} \, W/c^2$, $\beta = 0.01$, $B_0 = 100T$, and elliptical polarization state.

**Fig. 4.** Variations of magnitude of the normalized electric field as a function of normalized z coordinate over plasma length for different electron density and Gaussian profile, $\beta = 0.01$, $B_0 = 100T$, and elliptical polarization state.

**Fig. 5.** Effect of various external magnetic field on the magnitude of the normalized electric field as a function of normalized z coordinate over plasma length for Gaussian profile, $n = 3.5 \times 10^{16} \, W/c^2$, $\beta = 0.01$, and elliptical polarization state.

**Fig. 6.** Plot of the normalized electric field as a function of normalized z coordinate over plasma length for different polarization states and Gaussian profile, $n = 3.5 \times 10^{16} \, W/c^2$, $\beta = 0.01$, $B_0 = 100T$.

**Fig. 7.** Variations of magnitude of the normalized electric field as a function of normalized z coordinate over plasma length for different of $\beta$ parameter and Gaussian profile, $n = 3.5 \times 10^{16} \, W/c^2$, $B_0 = 100T$, and elliptical polarization state.

**Fig. 8.** Effect of various spatial laser profile on the magnitude of the normalized electric field as a function of normalized z coordinate over plasma length for $n = 3.5 \times 10^{16} \, W/c^2$, $\beta = 0.01$, $B_0 = 100T$, and elliptical polarization state.

**Fig. 8.** Variations of magnitude of the normalized intensity as a function of normalized retarded time throughout the duration of the laser pulse for different laser profiles, $n = 3.5 \times 10^{16} \, W/c^2$, $\beta = 0.01$, $B_0 = 100T$, and elliptical polarization state.

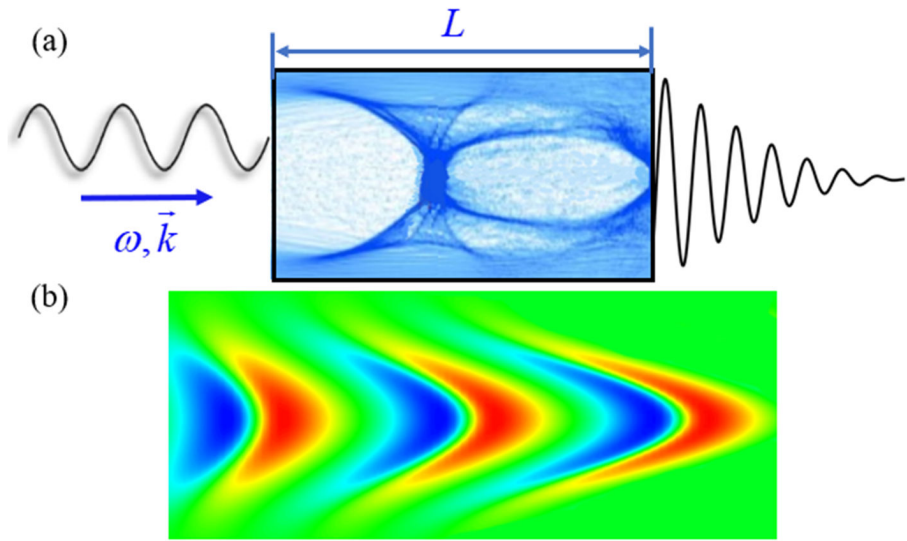

**Fig. 1.**

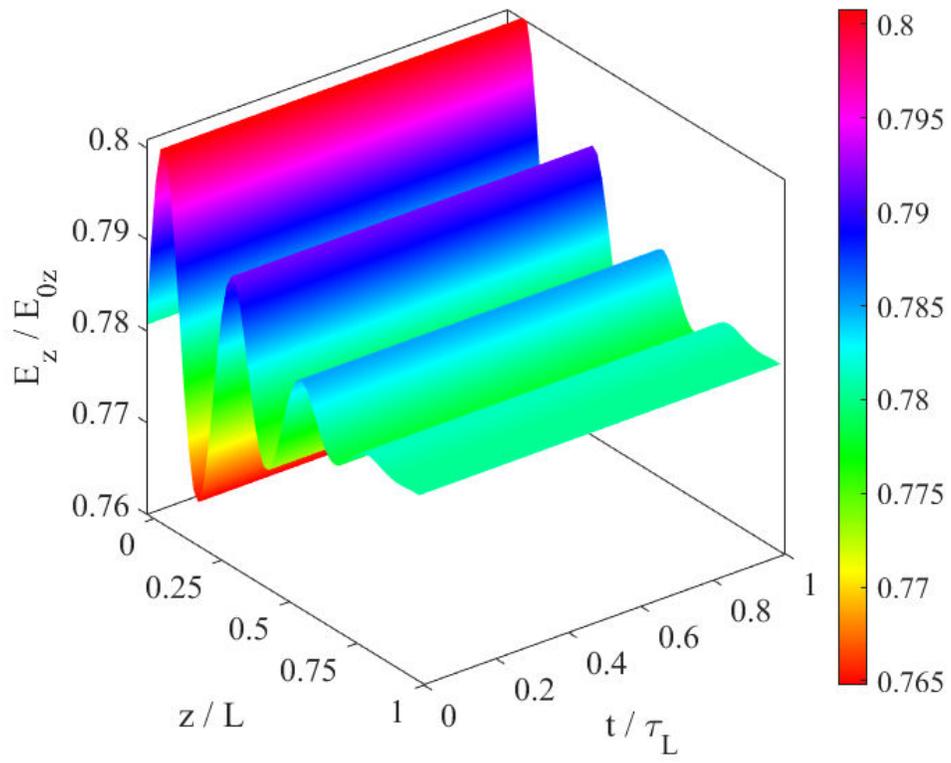

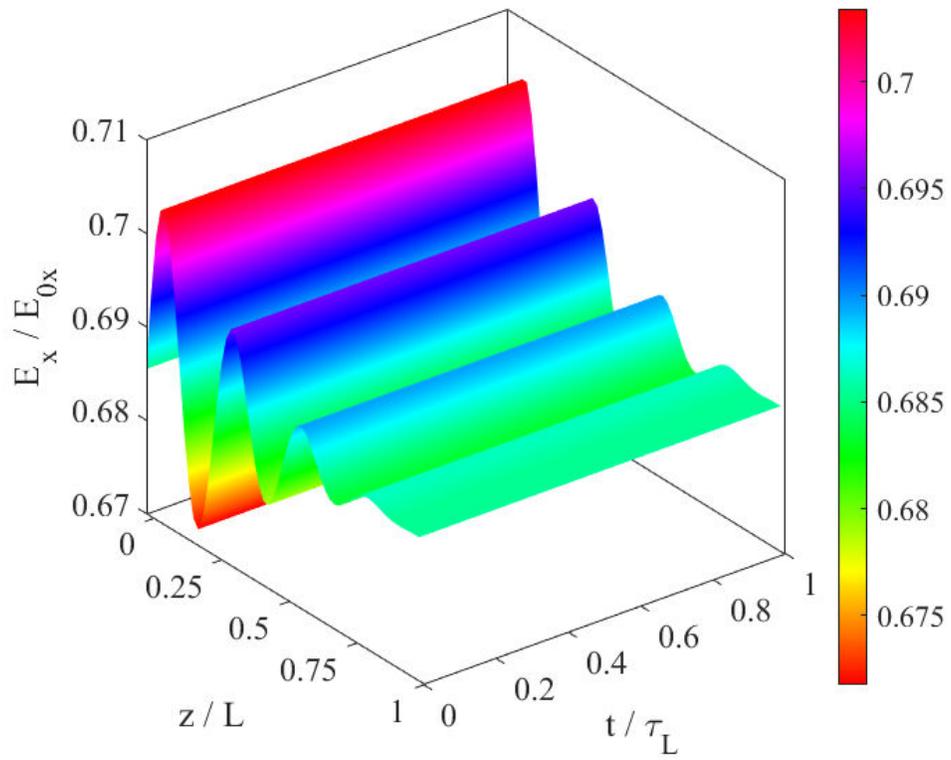

**Fig. 2.**

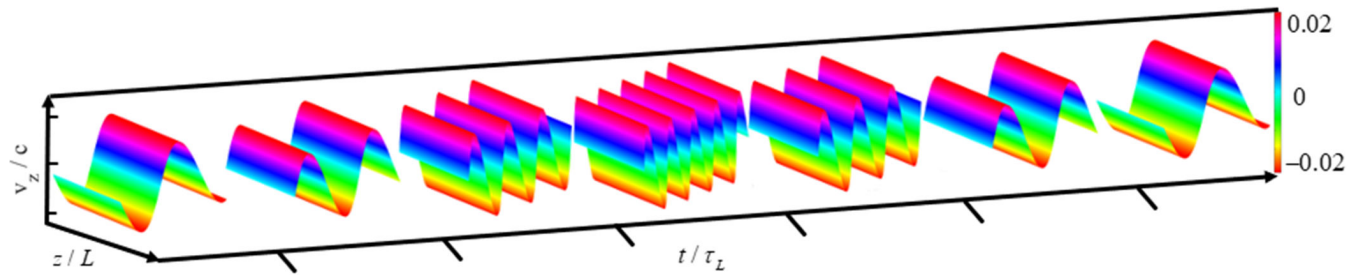

**Fig. 3.**

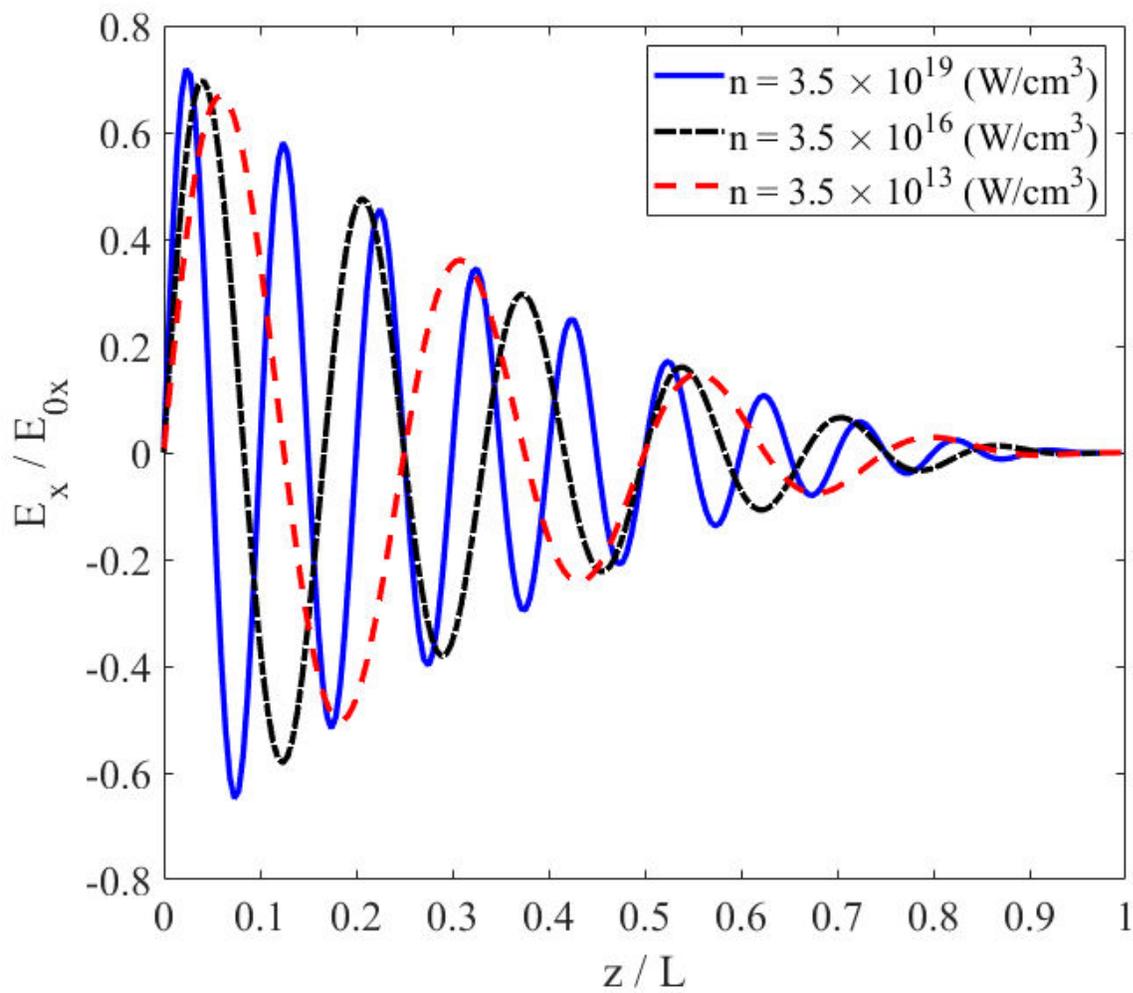

**Fig. 4.**

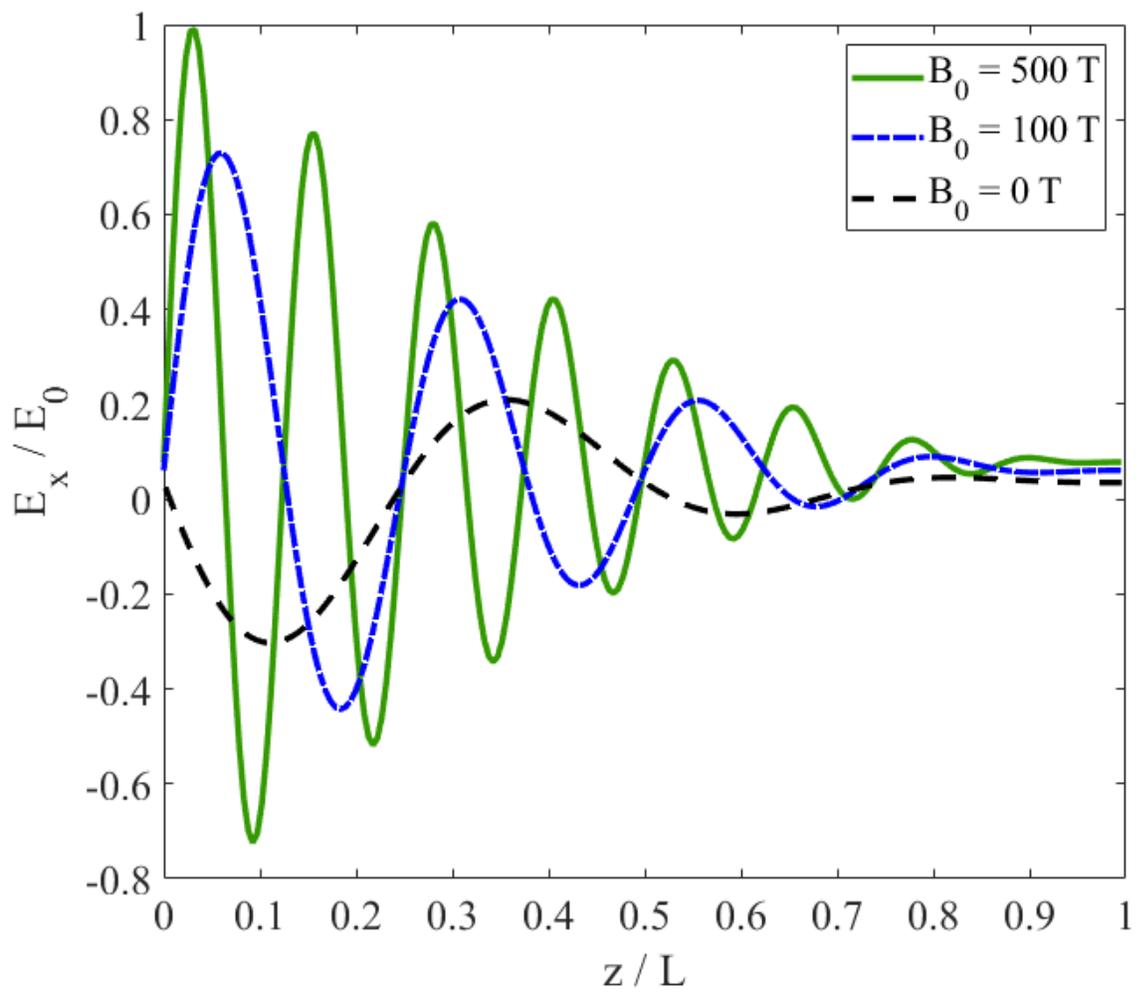

**Fig. 5.**

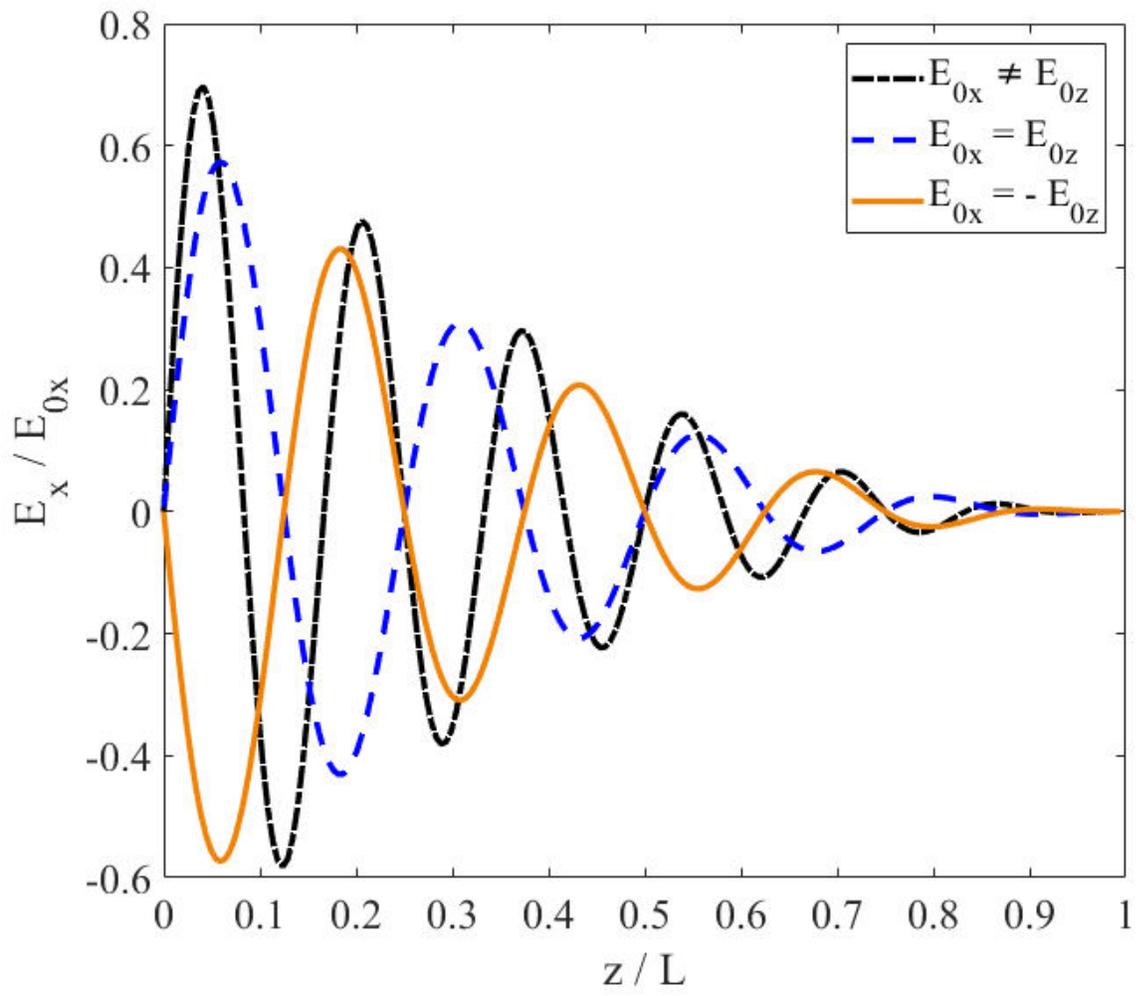

Fig. 6.

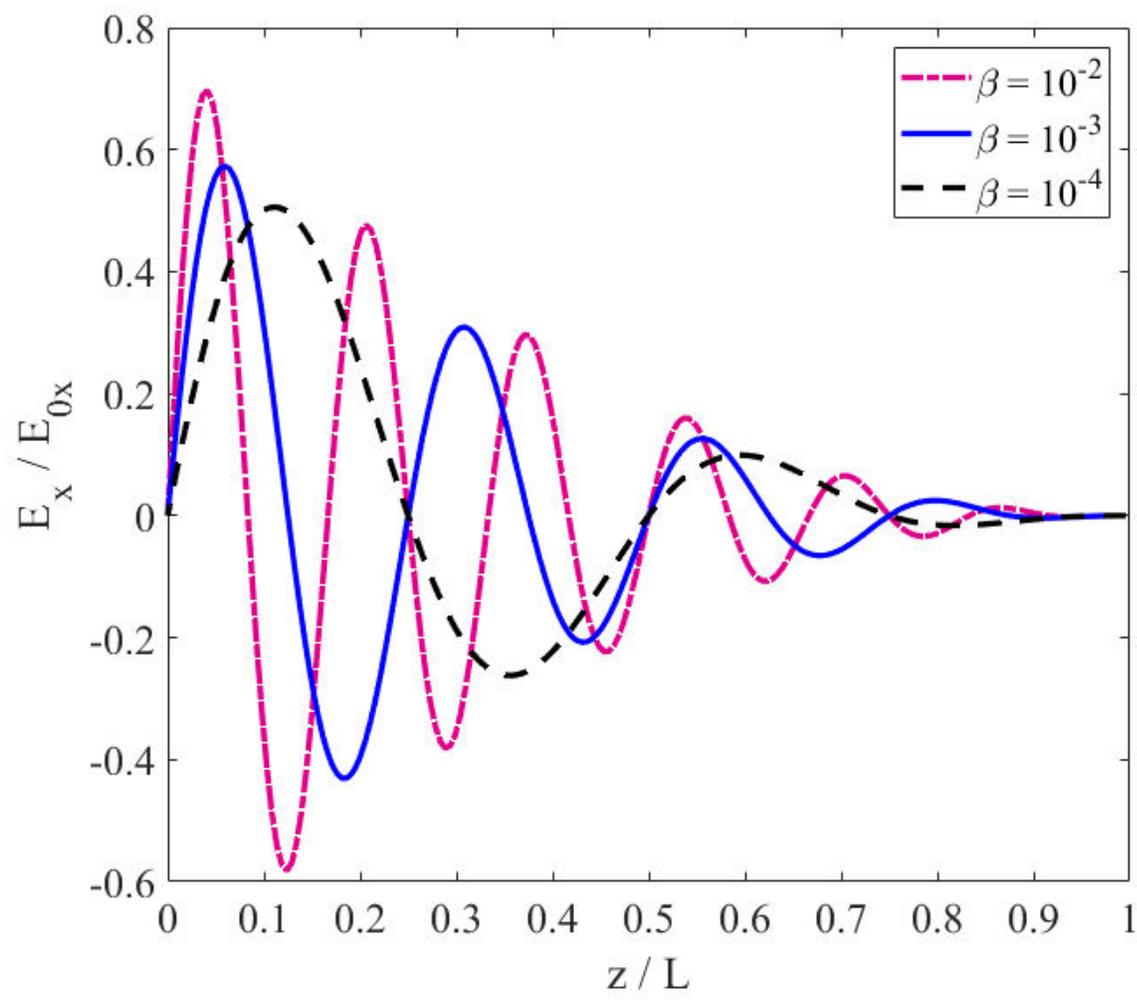

**Fig. 7.**

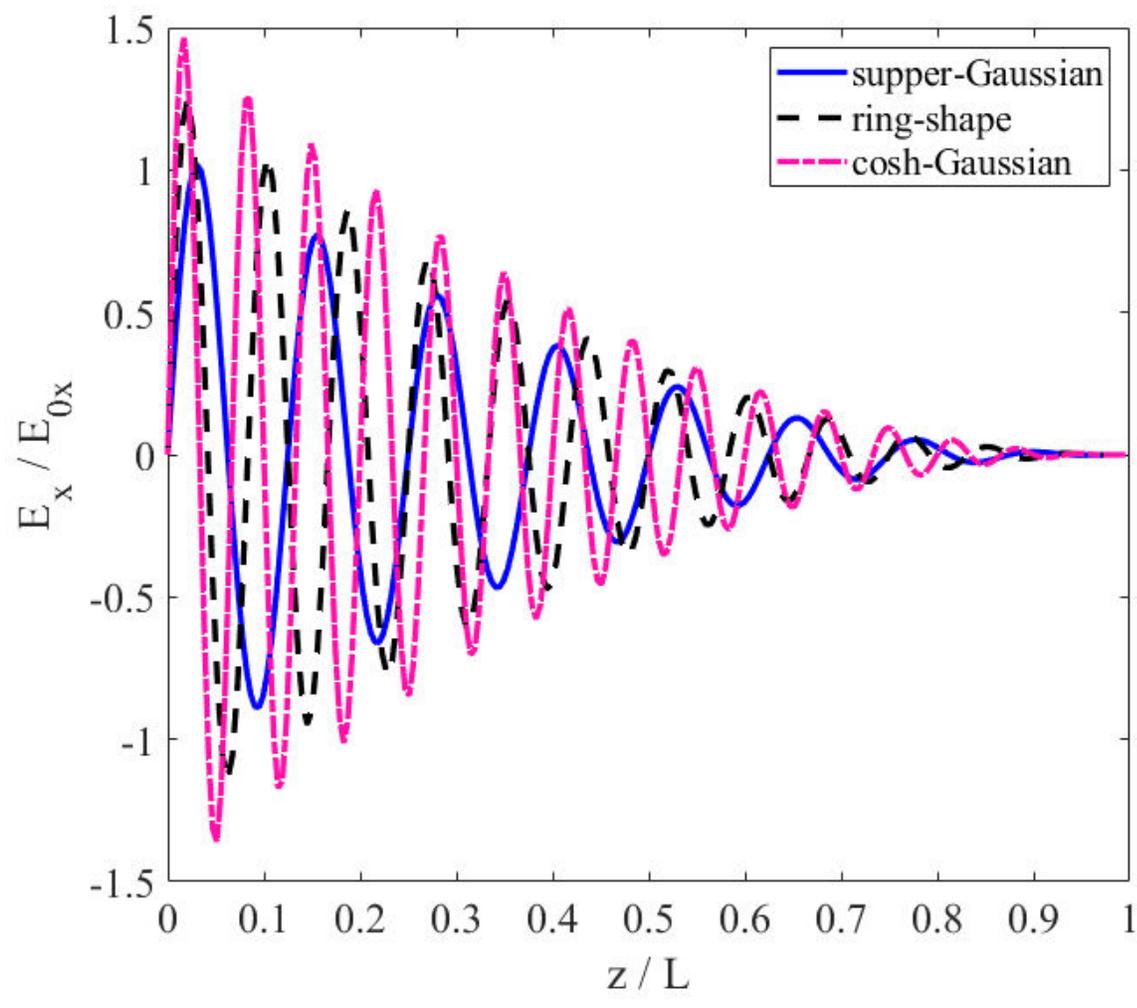

**Fig. 8.**

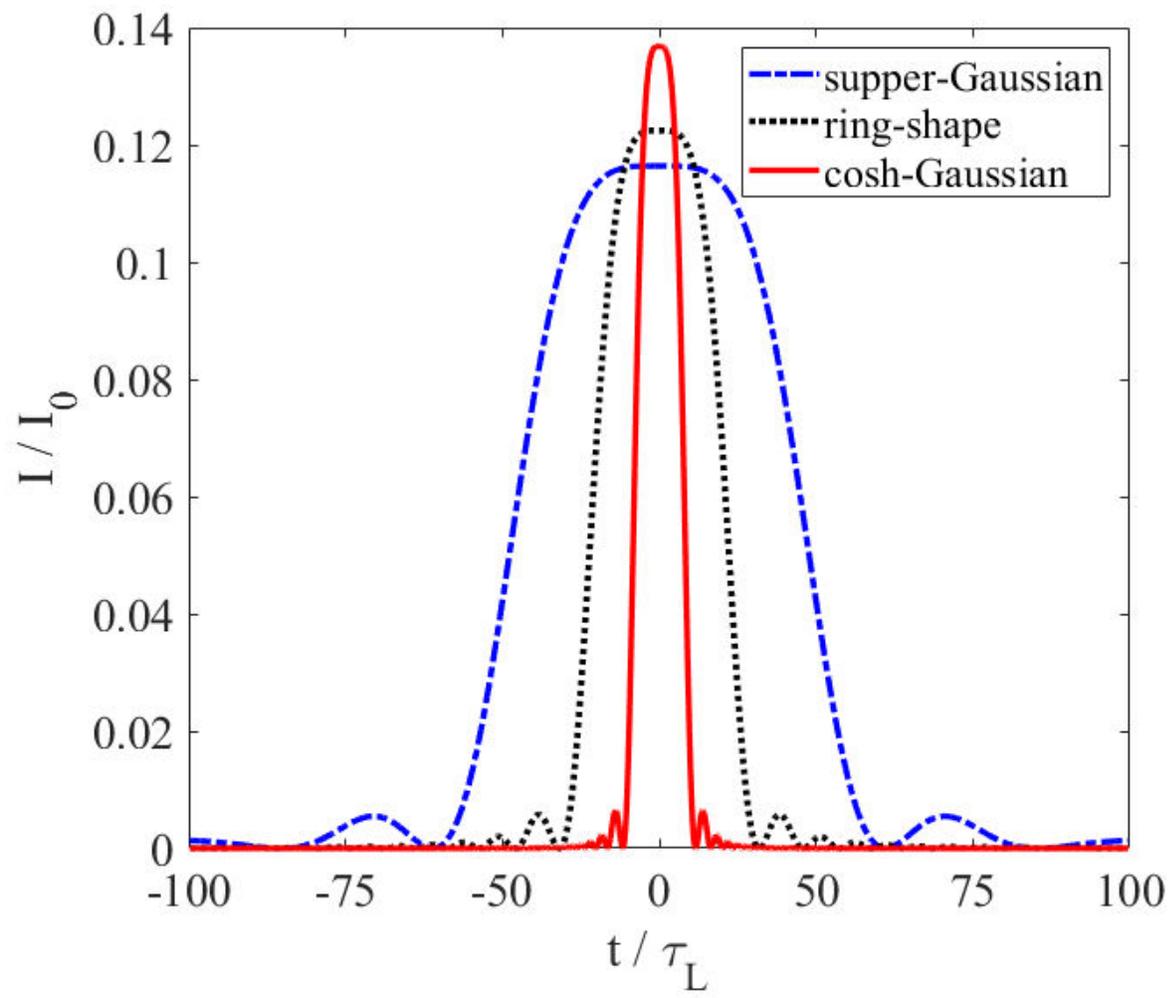

**Fig. 9.**